\documentclass{PHYEAUTH}
\usepackage{graphicx}
\usepackage{amsmath}
\usepackage{amssymb}

\begin{document}

\begin{frontmatter}

\title{Magnetoresistance oscillations in GaAs/AlGaAs superlattices
subject to in-plane magnetic fields }
\author[address1]{L.\ Smr\v{c}ka\thanksref{thank1}},
\author[address1]{P.\ Va\v{s}ek},
\author[address1]{P.\ Svoboda},
\author[address1]{N.\ A.\ Goncharuk},
\author[address1]{O.\ Pacherov\'{a}},
\author[address1,address2]{Yu.\ Krupko},
\author[address2]{Y.\ Sheikin}
and
\author[address3]{W.\ Wegscheider}
\address[address1]{Institute of Physics, Academy of Sciences of the
Czech Republic, Cukrovarnick\'a 10, 162 53 Prague 6, Czech Republic}
\address[address2]{Grenoble High Magnetic Field Laboratory, B.P. 166,
F-38042 Grenoble Cedex 9, France}
\address[address3]{Universit\"at Regensburg, Universit\"atstrasse 31,
D-93040 Regensburg, Germany}
\thanks[thank1]{
Corresponding author.
E-mail: smrcka@fzu.cz}
\begin{abstract}
The MBE-grown GaAs/AlGaAs superlattice with Si-doped barriers has been
used to study a 3D$\rightarrow$2D transition under the influence of
the in-plane component of applied magnetic field. The longitudinal
magnetoresistance data measured in tilted magnetic fields have been
interpreted in terms of a simple tight-binding model. The data provide
values of basic parameters of the model and make it possible to
reconstruct the superlattice Fermi surface and to calculate the
density of states for the lowest Landau subbands. Positions of van
Hove singularities in the DOS agree excellently with magnetoresistance
oscillations, confirming that the model describes adequately the
magnetoresistance of strongly coupled semiconductor superlattices.
\end{abstract}
\end{frontmatter}
%
\section{Introduction}
A superlattice (SL) is a set of regularly spaced quantum wells,
coupled together via tunneling across the barriers separating them.
In such a system, electrons move freely parallel to the plane of the
wells, while their perpendicular motion (i.e. that in the growth
direction) gives rise to the existence of energy minibands, whose
widths are proportional to the strength of inter-well coupling. In a
short-period SL, only the lowest miniband is usually occupied.

The quasiclassical interpretation of magnetotransport experiments
relies on the Onsager-Lifshitz quantization rule
\cite{on,Ashcroft}. For the longitudinal magnetoresistance, it leads
to Shubnikov-de Haas oscillations, which are periodic in $1/B$ and the
period is given by extremal cross-sections of the Fermi surface
perpendicular to the direction of the applied field $\vec{B}$. There
is, however, experimental evidence, that the quasiclassical
interpretation of the data may fail in semiconductor SLs in tilted
fields. The failure is attributed to the in-plane component of the
field $\vec{B}$, that suppresses electronic tunneling between wells,
when their separation becomes comparable with the in-plane magnetic
length $\ell_y = \sqrt{\hbar/|e|B_y}$~\,\cite{Dingle}.

We have recently studied the problem theoretically \cite{Goncharuk}.
The aim of this paper is to compare the longitudinal
magnetoresistance $\rho_{xx}(B)$, measured on the GaAs/AlGaAs
superlattice with rather strong inter-well coupling, with the
predictions of the above mentioned model calculations.
%
\begin{figure}[h,t,b]
\begin{center}\leavevmode
\includegraphics[width=0.9\linewidth]{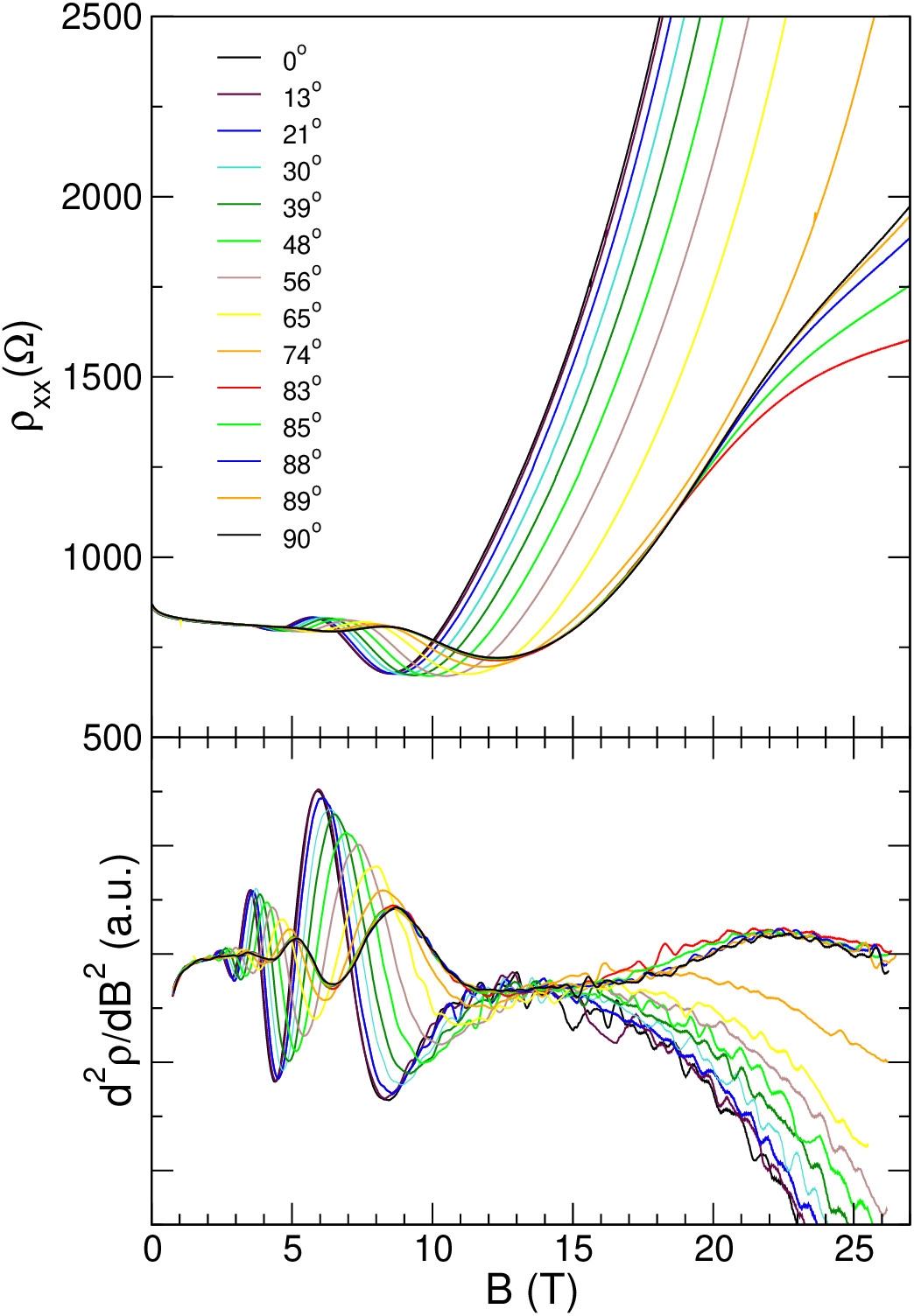}
\caption{Curves of the longitudinal magnetoresistance measured at 0.4
K for various tilt angles $\varphi$ are displayed in the upper part of
the figure. To emphasize the fine structure of the curves, their
second derivatives are shown in its lower part.}
\label{Fig1}
\end{center}
\end{figure}
%
\section{Experiments}
The MBE-grown SL with 30 periods consists of 29 GaAs quantum wells of
nominal width 5.0 nm, separated by $\rm Al_{0.3}Ga_{0.7}As$ barriers
of total nominal width 4.0 nm. Each barrier is composed of the inner
Si-doped layer 2.7 nm thick, surrounded by undoped 0.65 nm thick
spacers. The dopants provide electrons for 2D electron sheets in the
wells, but simultaneously limit their mobility and influence the
effective height of the barriers. Taking these nominal values, we get
for the SL period in the growth direction $d_z$ = 9.0 nm.

We have checked the periodicity of the structure using the X-ray
diffraction. It confirmed its excellent crystallographic quality, but
provided the slightly smaller period $d_z$ = 8.5 nm. We will use this
experimental value of $d_z$ below.

Several samples in the Hall bar geometry have been etched out of this
wafer and equipped with evaporated AuGeNi contacts. Both the
longitudinal and Hall resistances were measured at $T$ = 0.4 K in
magnetic fields up to 28 T. The sample could be rotated to any angle
between the perpendicular ($\varphi = 0^{\circ}$) and in-plane
($\varphi$ = 90$^{\circ}$) orientations. The standard low-frequency
($f$ = 13 Hz) lock-in technique has been used for the measurement.

Due to quasi-3D nature of the sample, it was not possible to determine
the effective electron concentration and mobility from the low-field
data.

In a weakly coupled system, deviating from the present one just by the
width of quantum wells (nominally about 19 nm instead of 5 nm) we
could observe well developed quantum Hall plateaus indicating that
27-28 2D layers of 29 were occupied with about the same electronic
concentration. We assume, that the same holds for the present sample
as well.
\section{Theory}
Let us shortly remind the essence of the model calculations presented
in Ref.\cite{Goncharuk}. In a tight-binding model of minibands in 3D
superlattices, we get an energy spectrum in the form
\begin{equation}
E(\vec{k}) = \frac{\hbar^2}{2m}(k_x^2+k_y^2) - 2t \cos(k_zd_z).
\label{energy}
\end{equation}
It consists of the energy of the free motion in the $x\,y$ plane and
the approximately cosine dispersion relation of the miniband in the
growth direction $z$. The parameter $t$ characterizes the effective
height and width of the barriers separating individual quantum
wells. The period $d_z$ of the SL determines the size of the Brillouin
zone: $ - \pi/d_z < k_z < \pi/d_z$. The Fermi surface of systems with
the Fermi energy $E_F$ within the miniband $(- 2t < E_F < 2t)$ has a
closed semielliptical shape. For $E_F > 2t$, it is an open corrugated
cylinder.

If we apply an external magnetic field $\vec{B}$, the energy spectrum
converts to a set of Landau subbands. Let us suppose, that $\vec{B}
\equiv ( 0, B\sin\varphi, B\cos\varphi)$, and that it changes its
orientation between perpendicular $(\varphi = 0^{\circ}, B = B_z)$ and
in-plane $(\varphi = 90^{\circ}, B = B_y)$ configurations.
Magneto-oscillations are determined by extremal cross-sections of the
Fermi surface and a plane perpendicular to $\vec{B}$. These
oscillations are periodic in $1/B$ with a period determined by the
cross-section area $A_k$~:
\begin{equation}
 A_k = \frac{2\pi |e|}{ \hbar}\frac{1}{\Delta (1/B)}.
\label{area}
\end{equation}
This quasiclassical approach works well at low magnetic fields. 

The above outlined description becomes completely inadequate for
strong in-plane fields $(B_z = 0)$. Matrix elements of the
one-electron Hamiltonian are then given in the tight-binding
approximation by
\begin{equation}
\label{H_sl_diagonal}
H_{j,j}=\frac{\hbar^2}{2m}\left(k_x+k_j\right)^2+
\frac{\hbar^2k_y^2}{2m}, \qquad H_{j,j\pm 1}=-t,
\end{equation}
where $k_j =j K_0$ is the magnetic-field-dependent wave-vector with
$K_0=|e|B_yd_z/\hbar=d_z/\ell_y^2$. The energy spectrum consists of
Landau subbands periodic in $k_x$ with the period $K_0$~:
\begin{equation}
E_n(k_x,k_y) = E_n(k_x) +\frac{\hbar^2k_y^2}{2m}.
\label{spec}
\end{equation}
The subbands $E_n(k_x)$ are flat and separated by wide gaps at low
$B_y$, at high fields the subbands become wide and the gaps narrow.
Therefore, the strong in-plane field $B_y$ changes the topology of the
equienergetic lines $E_F = E_n(k_x,k_y)$. For large $B_y$, $K_0$
becomes larger than the diameter of the free-electron Fermi circles
$2k_F$. Fermi contours do not cross anymore and inter-well tunneling
becomes impossible. This is the meaning of the statement, that strong
in-plane fields suppress inter-layer coupling and facilitate the
3D$\rightarrow$2D transition in SLs.

Upon lowering $B_y$, Fermi contours first touch at the Brillouin zone
boundaries, when $E_F$ reaches the top of the lowest Landau
subband. Then they merge into an open contour and new closed Fermi
ovals belonging to the next higher subband appear, when $E_F$ reaches
its bottom. These events correspond to two types of critical fields
$B_{c1}$ and $B_{c2}$.  At $B = B_{c1}$, $E_F$ touches the bottom of a
Landau subband. $E_n(k_x,k_y)$ has a minimum there and a step-like
singularity appears in the DOS. At $B= B_{c2}$, $E_F$ coincides with
the top of the subband, which corresponds to a saddle point in
$E_n(k_x,k_y)$ and to a logarithmic singularity in the DOS.
\begin{figure}[h,t,b]
\begin{center}\leavevmode
\includegraphics[width=0.9\linewidth]{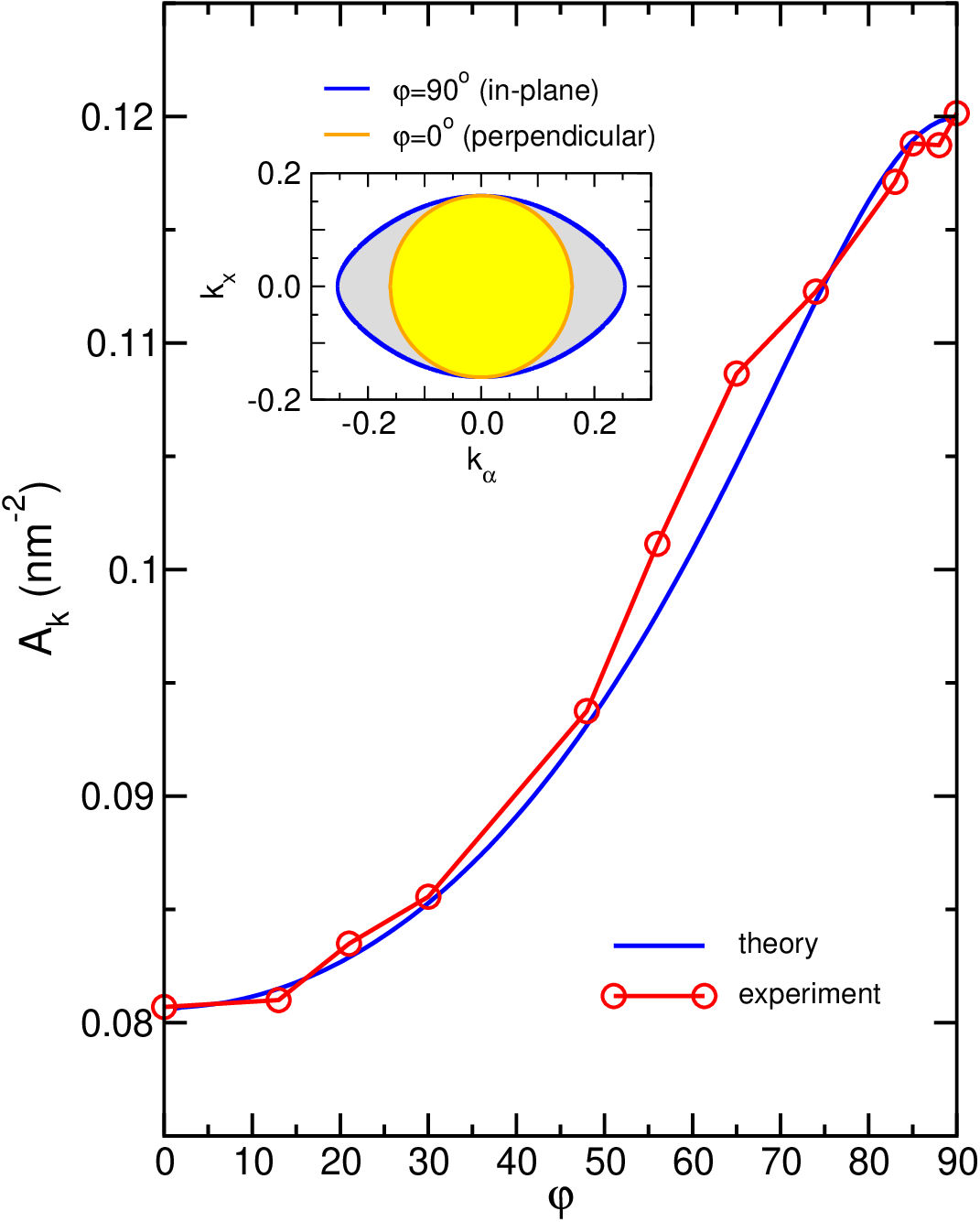}
\caption{The reconstruction of the Fermi surface from the
low-field-magnetoresistance data. The parameters $t$ = 4.7 meV and
$E_F$ = 5.2 meV lead to good agreement between the experimental and
theoretical curves. The inset shows two cross-sections of the Fermi
surface, the $k_{\alpha}$ on the horizontal axis denotes either $k_y$
or $k_z$. }
\label{Fig2}
\end{center}
\end{figure}
\section{Results and  discussion}
Experimental curves of the longitudinal magnetoresistance
$\rho_{xx}(B)$ for various tilt angles $\varphi$ are displayed in the
upper part of Fig.~\ref{Fig1}. To emphasize the fine structure of the
curves, their second derivatives are shown in its lower part. The
oscillations can be seen on all the curves, including that obtained in
strictly in-plane fields.  The oscillations of the magnetoresistance
in lower fields are found to be periodic in $1/B$ and the periods have
been employed to characterize the sample in terms of the model
described above.
\begin{figure}[h,t,b]
\begin{center}\leavevmode
\includegraphics[width=0.9\linewidth]{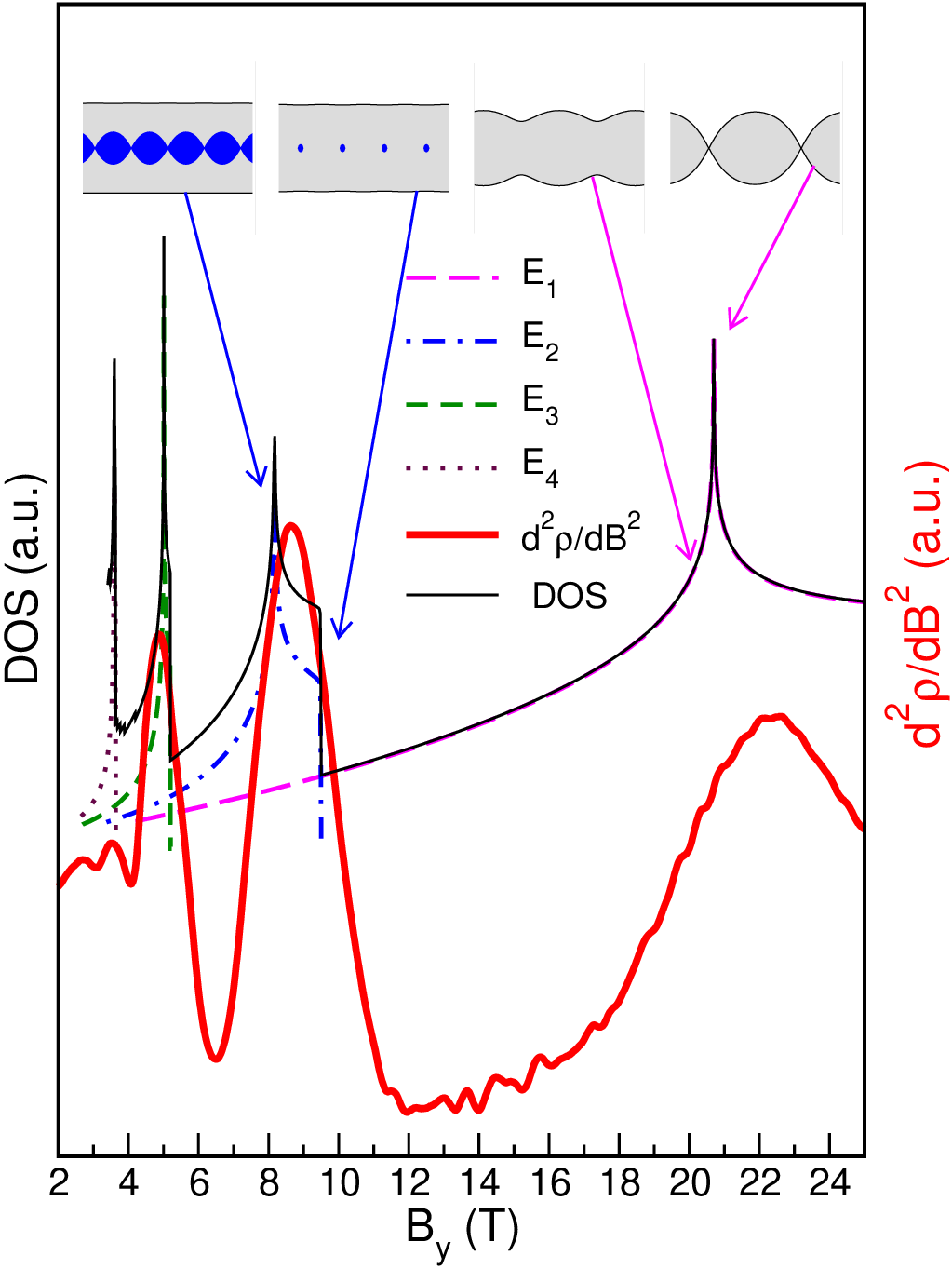}
\caption{The second derivative of the in-plane magnetoresistance
compared to DOS calculated as a function of $B_y$.}
\label{Fig3}
\end{center}
\end{figure}
Fitting the parameters of equation~(\ref{area}) to the experimental
data, we were able to get the values of the coupling constant $t$ =
4.7 meV and of the Fermi energy $E_F$ = 5.2 meV. The fit quality is
illustrated by Fig.~\ref{Fig2}. Experimental points stem from the
periods of oscillations. The parameters $t$ and $E_F$ have been
calculated from the data for two extreme tilt angles $\varphi =
0^{\circ}$ and $\varphi = 90^{\circ}$ and inserted to
equation~(\ref{area}) to get the theoretical interpolation curve drawn
in Fig.~\ref{Fig2}. The fit is quite good, confirming that the model
is adequate to describe the experimental data.

The data also allow to reconstruct the Fermi surface shape, as seen on
the inset of Fig.~\ref{Fig2}. As expected from the values of the model
parameters, it really has a form of a semielliptic closed body
entirely contained within the first Brillouin zone.

Experimentally determined parameters have been used to calculate the
DOS for four lowest Landau subbands. They are shown in Fig.~\ref{Fig3}
together with the resulting total DOS as functions of the in-plane
field $B_y$.  The logarithmic and step-like singularities can be
clearly distinguished in the $E_2$ and $E_3$ subbands only. In higher
subbands they merge and partial DOSs acquire the same form as that for
the strictly 3D systems. In the first subband only the logarithmic
singularity can exist. The pictures in the upper part of
Fig.~\ref{Fig3} illustrate changes of the Fermi surface topology
assigned to particular singularities in the DOS. Only the Fermi lines
of the lowest two Landau subbands are displayed.

The thick red curve in Fig.~\ref{Fig3} represents the second
derivative of the experimental magnetoresistance curve, corresponding
to the in-plane orientation of the applied magnetic field. There is
excellent agreement of the magnetoresistance peaks positions with the
maxima of the DOS. Slight asymmetry of the magnetoresistance peak
corresponding to the passing of $E_F$ through the subband $E_2$ may be
attributed to the separation of both DOS singularities due to the
finite width of the subband. The magnetoresistance maximum at the
highest field apparently reflects the logarithmic singularity in the
first Landau subband.
\section{Conclusions}
The periods of oscillations of low-field magnetoresistance were used
to construct the Fermi surface of the SL. The good fit of the
theoretical curve confirms that the two-parameter cosine model of the
miniband is appropriate to describe the experimental data.

The strong in-plane magnetic field suppresses inter-well coupling and
finally converts the 3D electronic system of the SL to an multiple 2D
layer system. The description of this process requires a fully
quantum-mechanical approach.  The oscillations of the calculated DOS
are in excellent agreement with oscillations of experimental
magnetoresistance curves. It proves, that the model is consistent and
can be used to explain qualitatively the magnetoresistance data on
strongly coupled semiconductor superlattices.
\section*{Acknowledgements}
This work has been supported by the Grant Agency of the ASCR under
Grant No. IAA1010408 and by the European Community contract
N$^{\circ}$ RITA-CT-2003-505474.


\begin{thebibliography}{9}
\bibitem{on} 
L.\ Onsager, Phil.\ Mag.\ {\bf 43}, 1006 (1952).
\bibitem{Ashcroft} 
N.\ W.\ Ashcroft and N.\ D.\ Mermin, in {\it Solid State Physics} 
(Saunders, Philadelphia, 1975), p.\ 268.
\bibitem{Dingle} 
R.\ Dingle, Surf.\ Sci.\ {\bf 73}, 229 (1978).
\bibitem{Goncharuk}
N.\ A.\ Goncharuk, L.\ Smr\v{c}ka, J.\ Ku\v{c}era and
K.\ V\'yborn\'y, Phys.\ Rev.\ B {\bf 71},  195318 (2005). 
\end{thebibliography}
\end{document}